\documentclass[a4paper]{panl}
\usepackage{cite}
\usepackage{wrapfig}
\usepackage{graphicx}
\usepackage{amssymb}
\usepackage{amsfonts}
\usepackage{amsmath}
\usepackage{longtable}
\usepackage{rotating}
\usepackage{lscape}
\usepackage{epsfig}
\usepackage{multirow}

\makeatletter
\newcommand{\mll}{m_{ll}}
\newcommand*{\rom}[1]{\expandafter\@slowromancap\romannumeral #1@}
\newcommand*\rot{\rotatebox{90}}
\makeatother

\originalTeX
\begin{document}
% Journal sections (see http://pkp.jinr.ru/index.php/PEPAN_LETTERS/about/editorialPolicies#focusAndScope)
\issuearea{Methods of physical experiment}
% or in Russian
%\issuearea{ФИЗИКА ЭЛЕМЕНТАРНЫХ ЧАСТИЦ И АТОМНОГО ЯДРА. ТЕОРИЯ}

\title{A new model test in high energy physics in frequentist and Bayesian statistical formalisms.\\ Проверка новой модели в физике высоких энергий в рамках частотного и Байесовского статистических формализмов.}
\maketitle
\authors{A.\,Kamenshchikov$^{a}$\footnote{E-mail: Andrey.Kamenshchikov@ihep.ru}}
\from{$^{a}$\,State Research Center of Russian Federation - Institute for High Energy Physics of National Research Center ``Kurchatov Institute'', 1, Nauki square, 142281, Protvino, Moscow region, Russian Federation}

%Abstract
\begin{abstract}
A problem of a new physical model test given observed experimental data is a typical one for modern experiments of high energy physics (HEP). A solution of the problem may be provided with two alternative statistical formalisms, namely frequentist and Bayesian, which are widely spread in contemporary HEP searches. A characteristic experimental situation is modeled from general considerations and both the approaches are utilized in order to test a new model. The results are juxtaposed, what demonstrates their consistency in this work. An effect of a systematic uncertainty treatment in the statistical analysis is also considered.

\vspace{0.2cm}

Проблема проверки новой модели с использованием экспериментальных данных является типичной для современных экспериментов в физике высоких энергий (ФВЭ). Решение такой проблемы может быть получено в рамках двух альтернативных статистических формализмов, а именно: частотного и Байесовского, имеющих широкое распространение в поисковых анализах ФВЭ. В данной работе из общих соображений смоделирована экспериментальная ситуация и произведена проверка новой модели ФВЭ с помощью обоих статистических подходов. Полученные результаты сопоставлены и демонстрируют взаимную совместимость. Рассмотрено влияние выбора способа включения систематической погрешности в статистический анализ на результат проверки новой модели.

\end{abstract}
\vspace*{6pt}

\noindent
PACS:  02.50.$-$r; 02.50.Cw; 02.50.Ng; 02.50.Tt; 02.70.$-$c; 02.70.Rr; 02.70.Tt; 02.70.Uu; 02.90.$+$p; 05.10.$-$a; 05.10.Ln; 05.90.$+$m%44.25.$+$f; 44.90.$+$c

%Introduction
\section{Introduction}
\label{sec:intro}

Test of Standard Model's extensions is one of the most popular directions in high energy physics (HEP) since the Higgs boson discovery on the LHC. Hence a problem of a new physical model test against observed data is a typical one for modern experiments of HEP. In a case no evidences of a new model are found, limits on the model's parameters are usually set. A solution of the problem may be provided with two alternative statistical formalisms, namely frequentist and Bayesian, which are widely spread in contemporary HEP searches. A characteristic experimental situation is modeled from general considerations as described in Sec.~\ref{sec:probstatement}. An application of frequentist formalism to the considered case is illustrated in Sec.~\ref{sec:freq}, while a solution in Bayesian paradigm is described in Sec.~\ref{sec:bayes}. The problem of a systematic source incorporation into an analysis is discussed in Sec.~\ref{sec:sys}.\par

%Introduction
\section{Problem statement}
\label{sec:probstatement}

Typical conditions, which an experimentalist usually deals with in HEP search analyses, generally comprise datasets both collected by an experiment (data hereinafter) and modeled with Monte Carlo (MC) methods. MC datasets are dedicated to both background and considered model's signal processes. A general structure of a dataset is represented by events, which are characterized by variates, e.g. missing transverse momentum in collider experiments. Such a dataset is named unbinned as opposed to a case of binned dataset, which consists of frequency distributions of those variates. Binned datasets possess an advantage of a simpler processing, especially in the latest HEP experiments with high multiplicity of recorded events. Therefore, an input of a statistical analysis typically includes so-called templates \--- frequency distributions of variates for all mentioned kinds of datasets.\par

 A characteristic statistical configuration of a physical search with a binned dataset and templates is modeled on the first step of this work. A case of a single observable variate, namely dileptonic invariant mass $m_{ll}$ (a basic observable quantity e.g. in $Z'$ search analyses), is considered here for a definiteness but may be generalized to a configuration with multiple variates. The background and signal processes' templates are modeled, using basic concepts of the distribution theory \cite{Kendall:1}. The RooFit toolkit \cite{RooFit} is used for these purposes, since it provides powerful and flexible instrumentalities for an implementation of probability density functions (p.d.f.s) and datasets' modeling.\par

The background is produced as a composition of the three components, which are marked as A, B and C and characterized by individual p.d.f.s and fractions. The crystal ball p.d.f. $ \mathrm{CB}(\mll|\bar{x}=91.5, \sigma=10, \alpha=-2.5, n=0.9)$ is used in order to model the background A for it resembles a general behavior of $Z$ boson peak and Drell-Yan (DY) tail. The background B modeling implements the complementary error function, constructing the p.d.f. $\frac{1}{2}\mathrm{errfc}(\frac{\mll-\bar{x}}{\sqrt{2}\sigma}|\bar{x}=150,\sigma=50)$. Such a parametrization fits for processes that are distributed almost uniformly up to some scale and tend to gradually decrease beyond that scale due to any statistical and/or physical reasons, e.g. $t\bar{t}$ on the Tevatron or the LHC. The background C is modeled with the exponential p.d.f. $\mathrm{exp}(\mll|\tau=-0.005)$ and serves as an approximation for gradually decreasing processes, e.g. fake leptons on the Tevatron or the LHC. The fractions of the backgrounds A, B and C are set to 0.9, 0.05 and 0.05 respectively. A similar background composition of the $Z$ and DY, the $t\bar{t}$ and the fake leptons processes happens to appear in search analyses on the Tevatron and the LHC. The background A possesses a suppressive dominance in the total composition because of its peak part; the composition is different on the right tail of the distribution, where a new physics search typically takes place. It is to be emphasized that the introduced tripartite background model is just an illustrative simplification and real experimental landscape generally comprises a richer diversity of background processes, what, nonetheless, doesn't affect the approaches discussed further.\par

A widespread case of a new model test is related to a resonance of an unknown mass ($m^{\mathrm{sig}}$). A width of such a resonance is often dominated by experimental resolution effects, which lead to noticeably wider shapes than those from a Breit-Wigner physics resonance width. Hence a signal may be modeled by the Gaussian p.d.f. $\mathrm{G}(\mll|m^{\mathrm{sig}},\sigma=0.05)$, where the $\sigma$ happens to be of a few percents order level. A several values of $m^{\mathrm{sig}}$ is usually tested during a search analysis so the range of $m^{\mathrm{sig}}\in\left[500, 2600\right]$ GeV is scanned with the step of $100$ GeV in this case for a definiteness.\par

A total expected number of the background events is set to $10^{3}$ for a mere definiteness. The data are generated by the instrumentality of the total background model p.d.f. in the extended likelihood formalism, which includes a Poissonian fluctuation of yields w.r.t. the complete expectation, provided by the defined configuration. The mentioned background processes p.d.f.s, the total background composition p.d.f. and the signal p.d.f. at $m^{\mathrm{sig}}=1$ TeV are presented on Fig.~\ref{pdfs}. The generated data are also shown.\par

%============================= Figure ================================
\begin{figure}[htpb]
\begin{center}
\includegraphics[width=127mm]{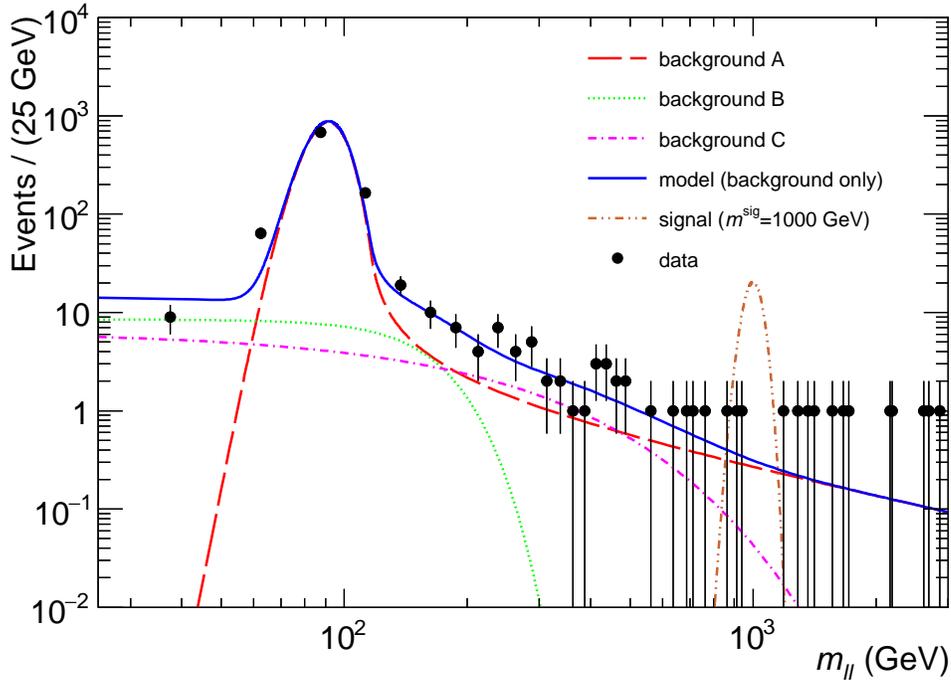} 
\vspace{-3mm}
\caption{P.d.f.s for the background and the signal processes and the data.}
\end{center}
\labelf{pdfs}
\vspace{-5mm}
\end{figure}

The MC samples of $10^{7}$ events for each of the background processes are generated and normalized according to the total background expectation and the predefined processes fractions. The signal MC samples are supplied with the $10^6$ events statistics for each considered $m^{\mathrm{sig}}$ and normalized to the nominal yield of 10 events just as a starting point: the signal yield is generally unknown and factorized to the nominal yield and the parameter of interest (POI) of an analysis $\mu$, signal strength. The defined configuration of the modeling leads to the distribution of the variate $m_{ll}$ on Fig.~\ref{dist}.\par

%============================= Figure ================================
\begin{figure}[htpb]
\begin{center}
\includegraphics[width=127mm]{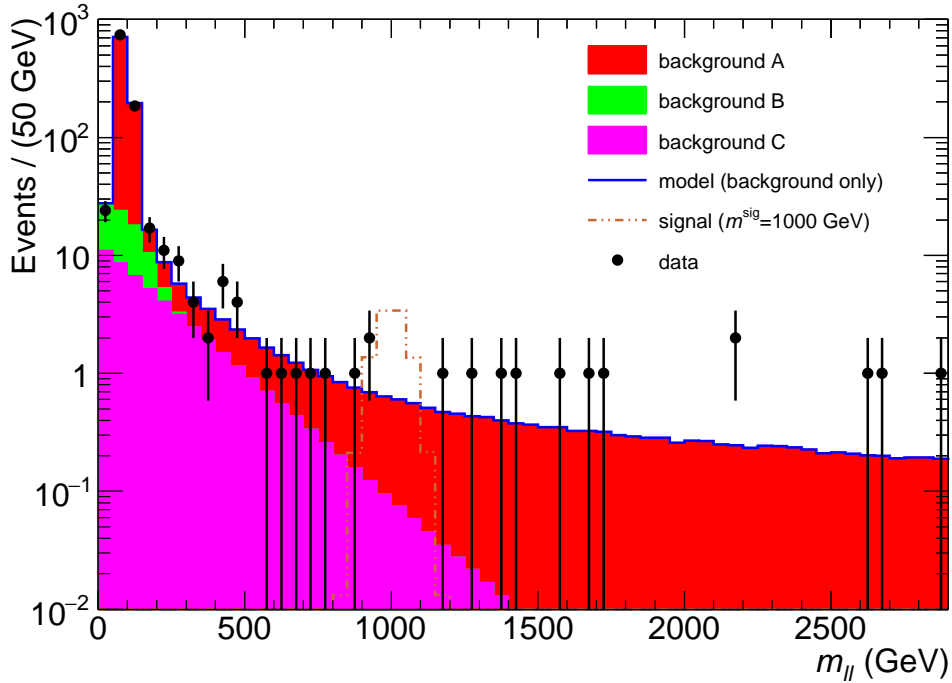} 
\vspace{-3mm}
\caption{Distributions of the $m_{ll}$ for the background and signal processes and the data.}
\end{center}
\labelf{dist}
\vspace{-5mm}
\end{figure}

The dedicated Control Region (CR) \--- an area of a phase space where a background component dominates w.r.t. the rest of the background \--- is defined in order to precise the background component's yield and to constrain its systematic variations. An area of a phase space, which is chosen to optimize a significance of a predicted signal appearance, is called Signal Region (SR). CR and SR concepts are actively used in modern physics searches. A natural choice of a CR in this case is an area around the background A peak, e.g. the interval $m_{ll}\in\left(60, 120\right)$ GeV, which guarantees an obvious dominance of the background A over the others. An SR should generally be optimized for each considered $m^{\mathrm{sig}}$, where a signal hypothesis is to be tested, and doesn't overlap with a CR. Therefore, a several SRs may be defined for this case. Taking into account the signal width, the SRs of this analysis can be set individually for each $m^{\mathrm{sig}}$ value under consideration: $m_{ll}>m^{\mathrm{sig}}-100$ GeV for $500\le m^{\mathrm{sig}}<1000$ GeV, $m_{ll}>m^{\mathrm{sig}}-200$ GeV for $1000\le m^{\mathrm{sig}}<1600$ GeV, $m_{ll}>m^{\mathrm{sig}}-300$ GeV for $1600\le m^{\mathrm{sig}}<2000$ GeV, $m_{ll}>m^{\mathrm{sig}}-400$ GeV for $2000\le m^{\mathrm{sig}}\le2600$ GeV. In a HEP analysis the CR and SR definition strategy is a subject of a detailed study and usually is an important part of a physics search. Here the numbers are defined from the very simple considerations and are chosen mostly for a definiteness, since this aspect is not a focus for this work. A simplest analysis configuration with single bin templates for CR and SRs \--- simple counting experiment \--- is discussed here and may be generalized to a case of multi-binned templates.\par

A common milestone of each search analysis is an evaluation and an implementation of a systematic uncertainty (s.u.). An s.u. may be considered as a variation of a systematic source (s.s.) and its impact on a yield estimate. An s.s. may be classified as the experimental, which comes from an inexactitude of physical quantities' estimates due to measuring or methodological imperfections, e.g. Jets Energy Scale (JES) variation, and the theoretical, which is from a lack of a theoretical knowledge, e.g. modeling uncertainty. An s.s.'s variation in a predefined direction leads to an impact of an individual size and direction for each separate process in every CR and SR of an analysis. A case of the same s.u.s configuration in all SRs is discussed here for a compactness. The next four s.s.s are introduced in the considered case: experimental s.s. \rom{1}, which affects backgrounds A and B and signal equally, e.g. like s.u. of luminosity; experimental s.s. \rom{2}, affects only background C, e.g. like the methodical s.u. on data-driven background; experimental s.s. \rom{3}, affects backgrounds A and B and signal individually, e.g. like JES s.u.; theoretical s.s. \rom{4}, affects backgrounds A and B, e.g. like s.u. of modeling. The detailed configuration of the s.u.s and s.s.s for, proposed for this case, is summarized in Tab.~\ref{Task:Systematics}.\par

\begin{table}[htpb]
%    \tiny
    \begin{center}
	\begin{tabular}{|l|c|c|c||c|c|c|c|}
	\hline
	\multirow{2}{*}{}    & \multicolumn{3}{|c||}{CR} & \multicolumn{4}{|c|}{SR}\\\cline{2-8}
				& \rot{Background A} & \rot{Background B} & \rot{Background C} & \rot{Background A} & \rot{Background B} & \rot{Background C} & \rot{Signal}\\\hline
	Source \rom{1}       & $^{\uparrow+5}_{\downarrow-5}$ & $^{\uparrow+5}_{\downarrow-5}$ & n/a & $^{\uparrow+5}_{\downarrow-5}$ & $^{\uparrow+5}_{\downarrow-5}$ & n/a & $^{\uparrow+5}_{\downarrow-5}$\\\hline
	Source \rom{2}       & n/a & n/a & $^{\uparrow+15}_{\downarrow-20}$ & n/a & n/a & $^{\uparrow-10}_{\downarrow+15}$ & n/a\\\hline
	Source \rom{3}       & $^{\uparrow+10}_{\downarrow-10}$ & $^{\uparrow+5}_{\downarrow-0}$ & n/a & $^{\uparrow-5}_{\downarrow+5}$ & $^{\uparrow+10}_{\downarrow-10}$ & n/a & $^{\uparrow+5}_{\downarrow-5}$\\\hline
	Source \rom{4}       & n/a & n/a & n/a & $^{\uparrow+20}_{\downarrow \mathrm{n/e}}$ & $^{\uparrow+10}_{\downarrow \mathrm{n/e}}$ & n/a & n/a\\\hline
	\end{tabular}
    \end{center}
\caption{The s.u.s and s.s.s configuration. Arrow -- a direction of a source's variation, signed numbers -- appropriate impacts in \%. ``n/a'' means that the s.s. doesn't affect the process. ``n/e'' indicates that the effect of corresponding s.s. variation in the respective direction is not estimated.}
\label{Task:Systematics}
\end{table}

A typical aim of a search analysis in HEP is to check a consistency of a background only model with an observation and to set upper limits on $\mu$. The parametrization of the processes and their composition, as well as the s.u.s, are chosen from the general considerations and the particular parameters' and variations' values are set for a definiteness. The idea of such a problem statement is to reproduce typical physical and statistical conditions of search experiments in HEP and to illustrate a solution of the problem in the frequentist and Bayesian statistical formalisms.\par

%Frequentist approach
\section{The frequentist approach}
\label{sec:freq}

A pivot of frequentist formalism is the likelihood function (LF) \cite{Kendall:2}, which generalizes all knowledge and understanding of an experiment, namely: observations, systematic variations, etc. The LF, which is also called model, is built up the next way for the introduced case:

\begin{equation}
	\begin{aligned}	
		\mathrm{L}(\boldsymbol{N},\boldsymbol{\theta^0},\boldsymbol{m}|\mu,\boldsymbol{\beta},\boldsymbol{\theta},\boldsymbol{\gamma})=\\
		\underbrace{\mathrm{P}\left(N_{\mathrm{SR}} \big|\overbrace{\left[\mu\times S_{\mathrm{SR}}\times\prod_{i}^{Sys} \mathrm{\nu}_{\mathrm{SR,sig}}^{i}\left(\theta_{i}\right)+\sum_{l}^{Bkg}\beta_{l}\times B_{\mathrm{SR},l}\times \prod_{i}^{Sys}\mathrm{\nu}_{\mathrm{SR},l}^{i}\left(\theta_{i}\right)\right]}^{\mathrm{\xi_{\mathrm{SR}}}}\times\gamma_{\mathrm{SR}}\right)}_{\mathrm{Poissonian\, term\, for\, SR}}\\
		\times\underbrace{\mathrm{P}\left(N_{\mathrm{CR}} \big|\overbrace{\left[\sum_{l}^{Bkg}\beta_{l}\times B_{\mathrm{CR},l}\times \prod_{i}^{Sys}\mathrm{\nu}_{\mathrm{CR},l}^{i}\left(\theta_{i}\right)\right]}^{\mathrm{\xi_{\mathrm{CR}}}}\times\gamma_{\mathrm{CR}}\right)}_{\mathrm{Poissonian\, term\, for\, CR}}\\
		\times\underbrace{\prod_{n}^{Sys} \mathrm{G}\left(\theta_{n}^0 \big| \theta_{n},1\right)}_{\mathrm{Gaussian\, constraint\, term}} \times \underbrace{\prod_{p}^{Reg} \mathrm{P}\left(m_{p} \big| \gamma_{p} \times \tau_{p}\right).}_{\mathrm{Poissonian\, constraint\, term}}
	\end{aligned}
	\label{Eq::Likelihood}
\end{equation}

The $Reg$ quantity is the set of the regions, defined in Sec.~\ref{sec:intro}, and the $\boldsymbol{N}$ numbers are the observed yields in those region. The s.u.s and the background components, described in Sec.~\ref{sec:intro}, are introduced via $Sys$ and $Bkg$ quantities. The renormalization parameters $\boldsymbol{\beta}$ are possessed by the backgrounds that are supplied with dedicated CRs (Background A) and float for them during maximization and integration procedures; they are merely set to unity $\beta\equiv1$ and fixed for the other backgrounds (Background B, Background C). The $B$ and $S$ numbers describe the predicted contributions of a background and a signal in a region. The impact function $\nu \left( \theta \right)$ introduces an effect of an s.s.'s variation to the model and is parametrized via a nuisance parameter (NP) $\theta$. NP variations are normally constrained by auxiliary measurements, which are expressed in the estimated yields under the various states of s.s.s. Since the Source \rom{1} affects all concerned processes with equal strength in CR and SRs, the dedicated impact function is built in a simplest manner as $\nu \left( \theta \right)=\theta$. For the remaining s.s.s a $\theta$ value of 0 conventionally corresponds to a nominal yield estimate ($I^{0}$), a value of unity corresponds to an estimate after a $1\sigma$ up variation of an s.s. ($I^{+}$), and a value of -1 corresponds to an estimate after a $1\sigma$ down variation of an s.s. ($I^{-}$). Following the procedure proposed for the LHC \cite{Cranmer} the polynomial interpolation and the exponential extrapolation is applied in order to construct the $\nu \left( \theta \right)$ with the $\theta\in(-\infty,+\infty)$:
\begin{equation}
	\mathrm{\nu}\left(\theta\big|I^{0},I^{+},I^{-}\right)=
	\begin{cases}
		\left(I^{+}/I^{0}\right)^{\theta} & \theta\ge1,\\
		1+\sum_{i=1}^{6}a_{i}\times\theta^{i} & \left|\theta\right|<1,\\
		\left(I^{-}/I^{0}\right)^{-\theta} & \theta\le-1.
	\end{cases}
	\label{Eq::Interpolation}
\end{equation}
In a case of only an up variation is available, e.g. Source \rom{4}, a down yield is taken symmetrically. The coefficients $a_{i}$ are calculated from the boundary conditions $\mathrm{\nu}\left(\theta=\pm1\right)$, $d\mathrm{\nu}/d\theta\big|_{\theta=\pm1}$, $d^2\mathrm{\nu}/d^2\theta\big|_{\theta=\pm1}$. This type of parametrization avoids kinks because of the continuous first and second derivatives and ensures that $\mathrm{\nu}\left( \theta \right)\ge0$ at any $\theta$. MC statistics limitedness of the processes' samples is accounted via the approach that is proposed for LHC \cite{Cranmer}: $\boldsymbol{\gamma}$ parameters introduce the effect in CR and SR and fluctuate around unity during maximization and integration procedures. $\boldsymbol{m}$ variates in the Poissonian constraint terms are defined as $m=\left(\xi/\delta\right)^2$, where $\xi$ is a total estimated yield in a region, subjected to the effect of MC sample limitedness, and $\delta$ is a total statistical uncertainty of that yield. If a yield is not subjected to the effect of MC statistics limitedness, corresponding terms are moved outside of the $\boldsymbol{\xi}$ sums in Eq.~\ref{Eq::Likelihood}, hence are not multiplied by the parameters $\boldsymbol{\gamma}$. The $\tau=\left(\xi/\delta\right)^2$ quantity is fixed in the model. The $\boldsymbol{\theta^0}$ and the $\boldsymbol{m}$ sets correspond to nominal yields estimates in the auxiliary measurements of an analysis, therefore, considering data, $\theta^0$ is set to 1 for the Source \rom{1} and to 0 for the remaining sources, while all $m$s are set to their initially estimated values. With such an approach the s.s. constraint terms of the model appear in a Gaussian form with the arguments $\boldsymbol{\theta^0}$, the mean $\boldsymbol{\theta}$ and the $\sigma$ is set to the value from Tab.~\ref{Task:Systematics} for the Source \rom{1} and to unity for the remaining sources. The $\boldsymbol{N}$ numbers are called observables whereas $\boldsymbol{\theta^0}$ and $\boldsymbol{m}$ are the global observables of an analysis. The $\boldsymbol{\beta}$, $\boldsymbol{\theta}$ and $\boldsymbol{\gamma}$ sets are named NPs.\par

The profile and projection for the POI and negative logarithmic likelihood (NLL) from Eq.~\ref{Eq::Likelihood}, offset by its global minimum value, at the $m^{\mathrm{sig}}=1000$ GeV are on Fig.~\ref{freqsol::poinllpll}.

%============================= Figure ================================
\begin{figure}[htpb]
\begin{center}
\includegraphics[width=127mm]{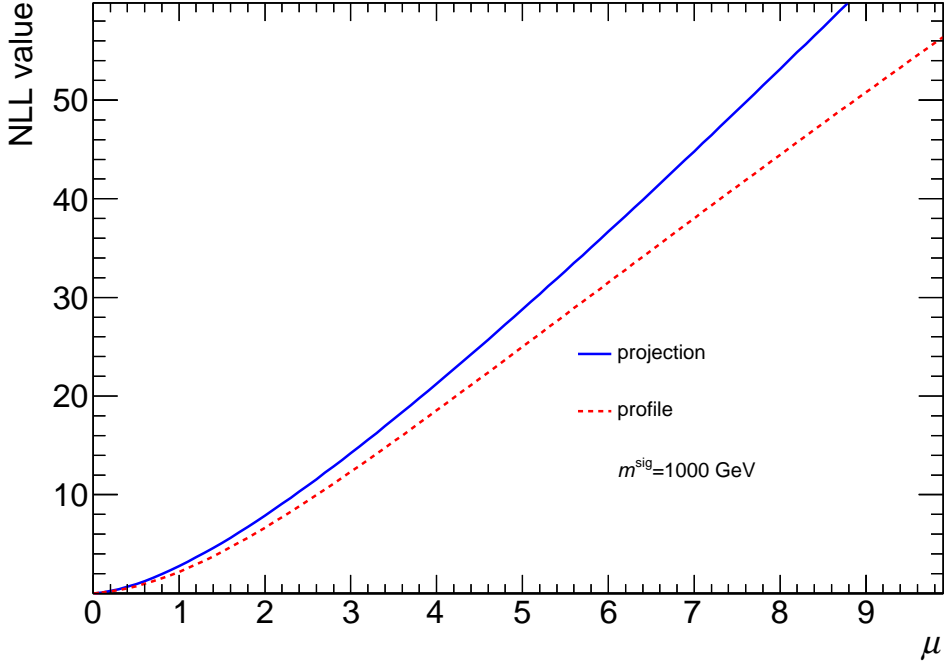} 
\vspace{-3mm}
\caption{The profile and projection for the POI and NLL at the $m^{\mathrm{sig}}=1000$ GeV.}
\end{center}
\labelf{freqsol::poinllpll}
\vspace{-5mm}
\end{figure}

The test statistic, which is based on the profile likelihood ratio, is used for the upper limit setting purpose following the study \cite{TestStatistics} as written in Eq.~\ref{Eq::qmu}.

\begin{equation}
	\tilde q_{\mu}=
	\begin{cases}
		-2 \ln \frac{\mathrm{L}\left(\mu,\mathrm{\hat{\hat{\boldsymbol{\beta}}}}\left(\mu\right),\mathrm{\hat{\hat{\boldsymbol{\theta}}}}\left(\mu\right),\mathrm{\hat{\hat{\boldsymbol{\gamma}}}}\left(\mu\right)\right)}{\mathrm{L}\left(0,\mathrm{\hat{\hat{\boldsymbol{\beta}}}}\left(0\right),\mathrm{\hat{\hat{\boldsymbol{\theta}}}}\left(0\right),\mathrm{\hat{\hat{\boldsymbol{\gamma}}}}\left(0\right)\right)} & \hat\mu<0,\\
		-2 \ln \frac{\mathrm{L}\left(\mu,\mathrm{\hat{\hat{\boldsymbol{\beta}}}}\left(\mu\right),\mathrm{\hat{\hat{\boldsymbol{\theta}}}}\left(\mu\right),\mathrm{\hat{\hat{\boldsymbol{\gamma}}}}\left(\mu\right)\right)}{\mathrm{L}\left(\hat\mu,\hat{\boldsymbol{\beta}},\hat{\boldsymbol{\theta}},\hat{\boldsymbol{\gamma}}\right)} & 0\le\hat\mu\le\mu,\\
		0 & \hat\mu>\mu.
	\end{cases}
	\label{Eq::qmu}
\end{equation}

A single hat symbol above a parameter means an unconditional maximization over that parameter on the whole its domain. A double hat symbol signifies the conditional maximization over that parameter given the fixed value of the parameter $\mu$ in parenthesis. Observables and global observables are considered as arguments of the model (Eq.~\ref{Eq::Likelihood}) and don't fluctuate during the maximization procedure. Since the test statistic of Eq.~\ref{Eq::qmu} rests on maximum likelihood (ML) estimates, the properties of consistency, unbiasedness and efficiency of ML estimator, some of which appear only asymptotically, should be considered for each particular case. The $p$ value is given by Eq.~\ref{Eq::pmu} in accordance with \cite{TestStatistics}:

\begin{equation}
	p_\mu=\int_{q_{\mu,obs}}^{\infty}\mathrm{f}\left(\tilde q_{\mu}\big|\mu\right)dq_\mu.
	\label{Eq::pmu}
\end{equation}

The p.d.f. $\mathrm{f}\left(\tilde q_{\mu}\big|\mu\right)$ is derived in a frequentist manner by the production of the pseudo-experiments (PEs). Unconditional ensembles are considered for this aim at each tested $\mu$ value: the $\boldsymbol{N}$, $\boldsymbol{\theta^0}$ and $\boldsymbol{m}$ sets do fluctuate during the production of the PEs according to the model in Eq.~\ref{Eq::Likelihood}, given the $\mathrm{\hat{\hat{\boldsymbol{\beta}}}}\left(\mu\right)$, $\mathrm{\hat{\hat{\boldsymbol{\theta}}}}\left(\mu\right)$ and $\mathrm{\hat{\hat{\boldsymbol{\gamma}}}}\left(\mu\right)$, extracted using maximization of the LF under a hypothesis of a given $\mu$ with the observables and global observables from data. The randomized quantities $\boldsymbol{N}$, $\boldsymbol{\theta^0}$ and $\boldsymbol{m}$ are treated as arguments for the LF in Eq.~\ref{Eq::Likelihood} which is eventually subjected to the maximization over the parameters as it is shown in Eq.~\ref{Eq::qmu} during each PE. Both signal+background ($s+b$) and background only ($b$) unconditional ensembles, which are of the sizes of 100000 PEs and 50000 PEs respectively, are produced at each $\mu$ point, providing the expected frequentist $\tilde q_{\mu}$ distributions for the both cases, as well as a single $\tilde q_{\mu}$ value for the data, which allows to get the $p$ values for the $s+b$ and $b$ hypotheses as defined in Eq.~\ref{Eq::pmu}. The HistFitter framework \cite{HistFitter}, which is based on the RooStats \cite{RooStats} classes, is utilized for a practical application of the frequentist approach to this case. The $\tilde q_{\mu}$ sampling distributions for the 20 POI values in the range of $\mu\in\left[0, 2.5\right]$ at the signal mass point $m^{\mathrm{sig}}=1000$ GeV are represented on Fig.~\ref{freqsol::stattestdist}.

%============================= Figure ================================
\begin{figure}[htpb]
\begin{center}
\includegraphics[width=127mm]{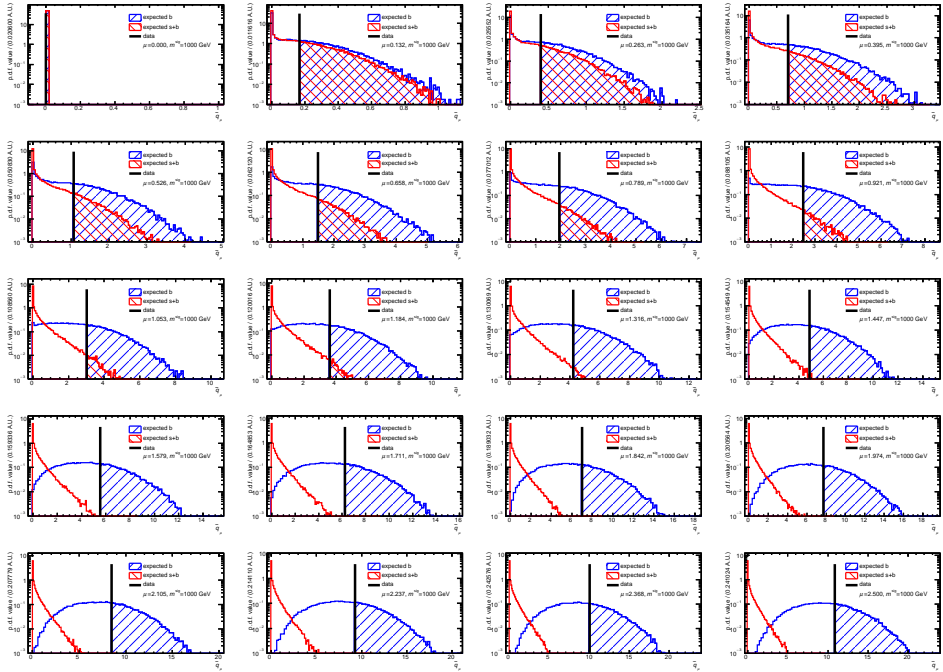} 
\vspace{-3mm}
\caption{$\tilde q_{\mu}$ distributions for the 20 scan points over $\mu\in\left[0, 2.5\right]$ at the $m^{\mathrm{sig}}=1000$ GeV.}
\end{center}
\labelf{freqsol::stattestdist}
\vspace{-5mm}
\end{figure}

The several observed $p$ values, which are also marked $CL$, are calculated at each $\mu$ point: $CL_{s+b}$, $CL_{b}$ and $CL_{s}$ where the latter one is introduced in \cite{CLs} and is written down in Eq.~\ref{Eq::CLs}:

\begin{equation}
	CL_{s}=\frac{CL_{s+b}}{1-CL_b}.
	\label{Eq::CLs}
\end{equation}

The $CL_s$ is known to be a conservative quantity, which generally leads to an overcoverage of an interval and hence looses a statistical sensitivity of an experiment. But nonetheless it is preferable in contemporary HEP analyses and is widely used in practice.\par

In a case that the observed $\tilde q_{\mu}$ value from data was substituted by the expected distribution of this quantity from an unconditional $b$ ensemble, the expected frequentist sampling distribution of the $CLs$ value for a case the signal doesn't exist becomes available, hence the median, the $\pm1\sigma$ band and the $\pm2\sigma$ band are merely the corresponding quantiles of that expected $CLs$ distribution. The observed and expected $CL$ values for the $m^{\mathrm{sig}}=1000$ GeV are on Fig.~\ref{freqsol::ulscan}. The observed and expected upper limits on the parameter $\mu$ along with the expected $\pm1\sigma$ and $\pm2\sigma$ values for the given $m^{\mathrm{sig}}$ follow from this scan plot just as the abscissas of the points of intersections of the respective $CLs$ curves with a Confidence Level (CL) set to be used for the reporting of an analysis's result: $95\%$ CL, which correspond to the $p$ value of 0.05, is typically used in searches of a new physics in HEP.\par

%============================= Figure ================================
\begin{figure}[htpb]
\begin{center}
\includegraphics[width=127mm]{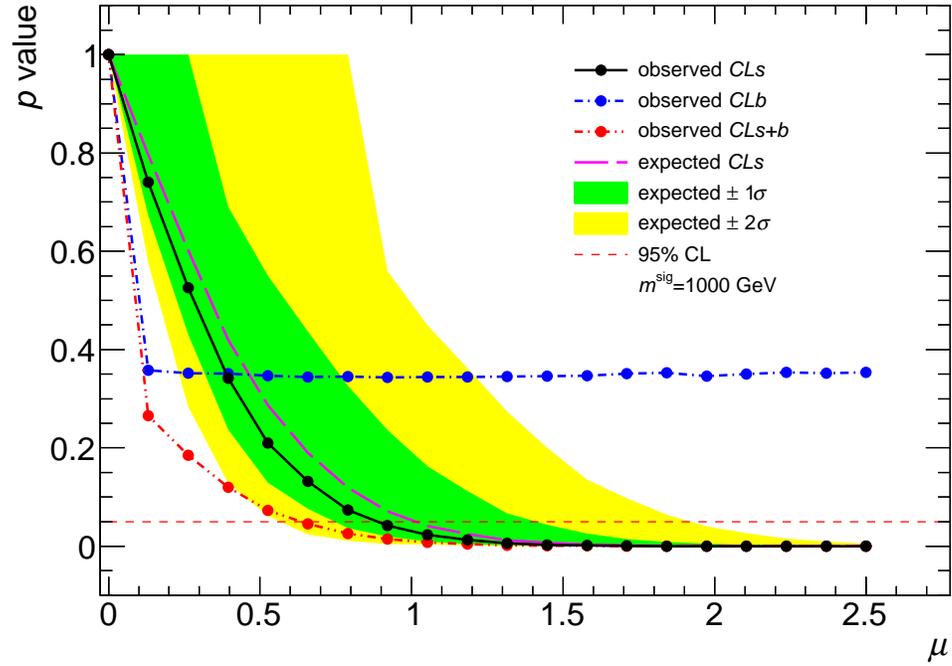} 
\vspace{-3mm}
\caption{$CL$ quantities for the 20 scan points over $\mu\in\left[0, 2.5\right]$ and $m^{\mathrm{sig}}=1000$ GeV.}
\end{center}
\labelf{freqsol::ulscan}
\vspace{-5mm}
\end{figure}

An application of such an approach for various $m^{\mathrm{sig}}$ values allows to build the observed and expected limits on the parameter $\mu$ as a function of $m^{\mathrm{sig}}$. A straightforward multiplication of those limits by the nominal signal samples yields at each $m^{\mathrm{sig}}$ point (10 events in this case) provides an interpretation of the results as the upper limits on the number of signal events ($N^{\mathrm{sig}}$) as a function of $m^{\mathrm{sig}}$ as it is shown on Fig.~\ref{freqsol::excl}.\par

%============================= Figure ================================
\begin{figure}[htpb]
\begin{center}
\includegraphics[width=127mm]{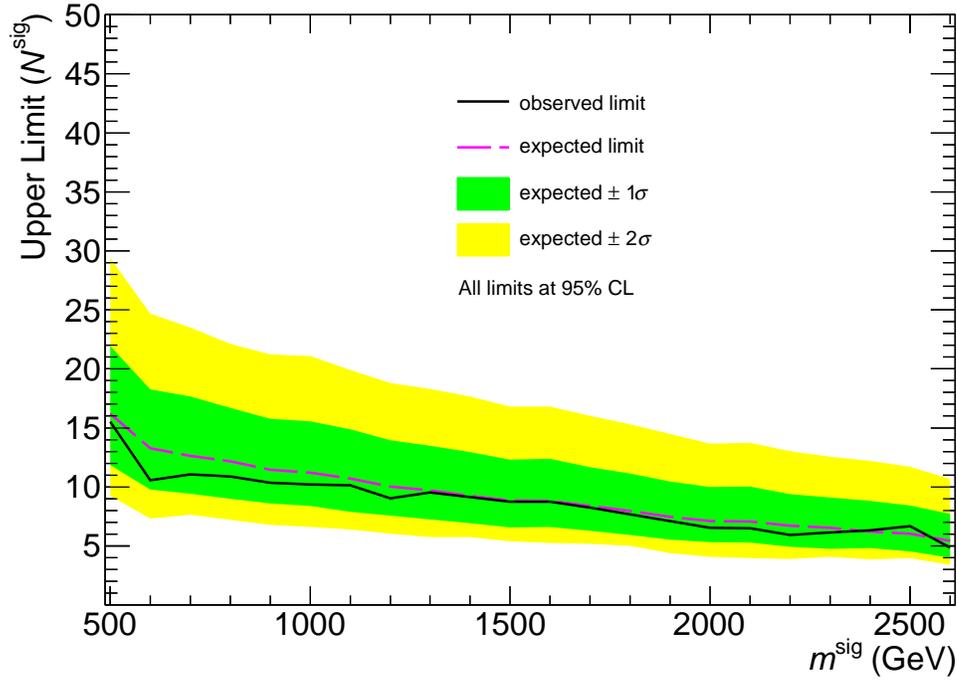}  
\vspace{-3mm}
\caption{Frequentist upper limits on $N^{\mathrm{sig}}$ as a function of $m^{\mathrm{sig}}$.}
\end{center}
\labelf{freqsol::excl}
\vspace{-5mm}
\end{figure}

A general tendency of the downward fluctuations (a deficit of the observed background events w.r.t. its nominal expectation) takes place for all tested signal mass points, leading to negative $\hat\mu$ values, except the $m^{\mathrm{sig}}=2500$ GeV point, where an opposite (upward) fluctuation presences. The effect is noticeable on Fig.~\ref{freqsol::excl} by a comparison of the observed limit and the expected limit polygons.

A feature of the modern search papers is to report the $p_{0}$ values. As it is proposed in \cite{TestStatistics}, the $q_0$ test statistic, which is used in order to get $p_0$ value, is not just a special case of $\tilde q_{\mu}$ in Eq.~\ref{Eq::qmu} and is defined in a different way as it is shown in Eq.~\ref{Eq::q0}:\par

\begin{equation}
	q_{0}=
	\begin{cases}
		-2 \ln \frac{\mathrm{L}\left(0,\mathrm{\hat{\hat{\boldsymbol{\beta}}}}\left(0\right),\mathrm{\hat{\hat{\boldsymbol{\theta}}}}\left(0\right),\mathrm{\hat{\hat{\boldsymbol{\gamma}}}}\left(0\right)\right)}{\mathrm{L}\left(\hat\mu,\hat{\boldsymbol{\beta}},\hat{\boldsymbol{\theta}},\hat{\boldsymbol{\gamma}}\right)} & \hat\mu\ge0,\\
		0 & \hat\mu<0.
	\end{cases}
	\label{Eq::q0}
\end{equation}

The rest of the procedure is the same as for $\tilde q_{\mu}$ case - the p.d.f. $\mathrm{f}(q_0\big|0)$ is derived in a frequentist manner by the instrumentality of 200000 $b$ PEs. $p_0$ value definition is written down in Eq.~\ref{Eq::p0}, according to \cite{TestStatistics}:\par

\begin{equation}
	p_0=\int_{q_{0,obs}}^{\infty}\mathrm{f}(q_0\big|0)dq_0.
	\label{Eq::p0}
\end{equation}

$p_0$ values as a function of $m^{\mathrm{sig}}$ are represented on Fig.~\ref{freqsol::disc}.\par

%============================= Figure ================================
\begin{figure}[htpb]
\begin{center}
\includegraphics[width=127mm]{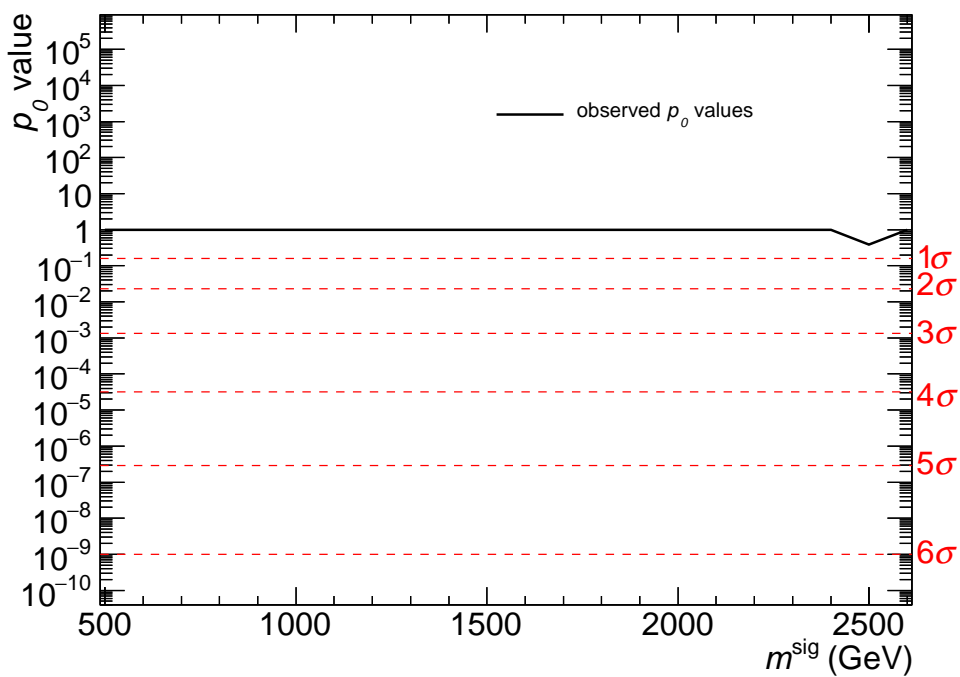} 
\vspace{-3mm}
\caption{Frequentist $p_0$ values as a function of $m^{\mathrm{sig}}$.}
\end{center}
\labelf{freqsol::disc}
\vspace{-5mm}
\end{figure}

The $p_0$ values are sticked to unity for the majority of mass points due to the downward fluctuations, which are not interpreted as a deviation from the background model by definition of $q_0$ test statistic. The spike at the $m^{\mathrm{sig}}=2500$ GeV point comes from an upward fluctuation which is quantified in terms of significance of a deviation from the background model ($<1\sigma$ for this case).

%Systematics treatment
\section{Systematic source incorporation}
\label{sec:sys}

An important point of the model construction is a treatment to an s.s. implementation in the LF in Eq.~\ref{Eq::Likelihood} \cite{Cranmer, Wanke:lla, Sivia:2013hca}. In particular the theoretical s.s. (Source \rom{4}) may be considered in a unique manner \cite{Cranmer, Heinrich:2007zza}, since it is generally based on a lack of theoretical knowledge \cite{Diehl:mla} rather than on auxiliary measurement estimate's interval as it typically happens with the majority of the experimental s.s.s. Hence a situation with a several independent yield's estimates at hand without any preconceptions and preferences about them may be encountered. A case with independent predicted background yields in an SR from different MC generators is an example of such an s.s. The Gaussian constraint term in the model of Eq.~\ref{Eq::Likelihood} may be considered to be replaced by a uniform term with the domain of $\theta\in\left( 0,1 \right)$. A global observable $\theta^0$ is absent for such an s.s. since it is not related with a measurement and introduces a freedom of the predicted background yield variation in an SR due to uncertain theoretical knowledge. The effect of this rearrangement is represented on the rebuilt upper limits on $N^{\mathrm{sig}}$ as a function of $m^{\mathrm{sig}}$ on Fig.~\ref{freqsol::constr}.\par

%============================= Figure ================================
\begin{figure}[htpb]
\begin{center}
\includegraphics[width=127mm]{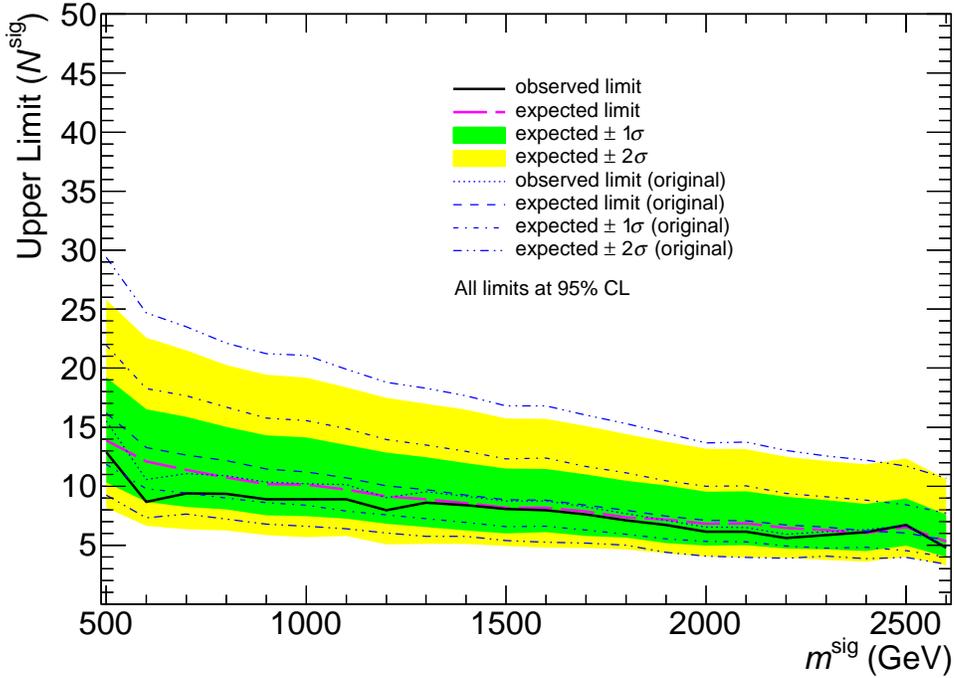} 
\vspace{-3mm}
\caption{Frequentist upper limits on $N^{\mathrm{sig}}$ as a function of $m^{\mathrm{sig}}$ with the uniform constraint term for Source \rom{4}.}
\end{center}
\labelf{freqsol::constr}
\vspace{-5mm}
\end{figure}

The original upper limits from Fig.~\ref{freqsol::excl} are superimposed on top of the  Fig.~\ref{freqsol::constr}. The new limits are stronger than the original ones because the Source \rom{4}'s NP variation is now restricted by the domain of the uniform term as opposed to the corresponding Gaussian term. In the same time, the observed and expected limits agree better at $m^{\mathrm{sig}}=2500$ GeV because of an improved flexibility of the background model in the domain of the Source \rom{4}'s NP.\par

It is to be stressed here for a completeness that multitude of possibilities may be considered for constraint term choice in LF \cite{Cranmer}, as well as various approaches can be applied to simplify the analysis of  LF with NPs \cite{Heinrich:2007zza}, since s.s. incorporation is generally subjective aspect and no unique recipe is available. The recommendation is to keep this points clear and well documented in HEP publications.

%Bayesian approach
\section{The Bayesian approach}
\label{sec:bayes}

Bayesian formalism, which rests on Bayes theorem, was being intensively developed for the second half of the twentieth century \cite{Kendall:3} and has gotten a wide spread in HEP analyses generally \cite{Schott:gla} and in cosmology particularly \cite{Prosper:mla, Trotta:2008qt}. Being applied to the considered situation, Bayes theorem allows us to get posterior p.d.f. $\mathrm{P}$ of analysis's parameters, both POI ($\mu$) and nuisance parameters ($\boldsymbol{\beta}, \boldsymbol{\theta},\boldsymbol{\gamma}$), given data ($\boldsymbol{N},\boldsymbol{\theta^0},\boldsymbol{m}$) by means of the LF $\mathrm{L_B}$ and the prior p.d.f. of the nuisance parameters $\mathrm{P_0}$ as it is written in Eq.~\ref{Eq::BayesTheorem}:
\begin{equation}
	\mathrm{P}(\mu,\boldsymbol{\beta},\boldsymbol{\theta},\boldsymbol{\gamma}|\boldsymbol{N},\boldsymbol{\theta^0},\boldsymbol{m})=\frac{\mathrm{L_B}(\boldsymbol{N},\boldsymbol{\theta^0},\boldsymbol{m}|\mu,\boldsymbol{\beta},\boldsymbol{\theta},\boldsymbol{\gamma})\times \mathrm{P_0}(\boldsymbol{\beta},\boldsymbol{\theta},\boldsymbol{\gamma}|\boldsymbol{\theta^0},\boldsymbol{m})}{\int \mathrm{L_B}(\boldsymbol{N},\boldsymbol{\theta^0},\boldsymbol{m}|\mu,\boldsymbol{\beta},\boldsymbol{\theta},\boldsymbol{\gamma})\times \mathrm{P_0}(\boldsymbol{\beta},\boldsymbol{\theta},\boldsymbol{\gamma}|\boldsymbol{\theta^0},\boldsymbol{m})d\mu d\boldsymbol{\beta}d\boldsymbol{\theta}d\boldsymbol{\gamma}}
	\label{Eq::BayesTheorem}
\end{equation}
The index in $\mathrm{L_B}$ is to emphasize the difference with respect to $\mathrm{L}$ in Eq.~\ref{Eq::Likelihood}, since the LF in Bayesian sense here comprises only the two former poissonian terms of Eq.~\ref{Eq::Likelihood}, given its reminder, which is a product of constraint terms in frequentist formalism, becomes the $\mathrm{P_0}(\boldsymbol{\beta},\boldsymbol{\theta},\boldsymbol{\gamma}|\boldsymbol{\theta^0},\boldsymbol{m})$ after a corresponding normalization. Hence the numerator of Eq.~\ref{Eq::BayesTheorem} and the likelihood in Eq.~\ref{Eq::Likelihood} are technically equivalent. An integration of the posterior p.d.f. $\mathrm{P}$ over the nuisance parameters ($\boldsymbol{\beta}, \boldsymbol{\theta},\boldsymbol{\gamma}$) is called marginalization and returns a posterior p.d.f. of the POI $\mathrm{P}(\mu|\boldsymbol{N},\boldsymbol{\theta^0},\boldsymbol{m})$ that allows to extract an upper limit on $\mu$ straightforwardly as a declared percentile ($0.95$ in this case) of the $\mathrm{P}(\mu)$. A part of POI domain below that percentile is called Credible Interval (CI).\par

The Bayesian Analysis Toolkit (BAT) \cite{BAT} is used as a framework for the Bayesian analysis where the marginalization process is provided with Markov Chains Monte Carlo (MCMC) techniques. The RooStats \cite{RooStats} classes are widely used during all operations. The $\mathrm{P}(\mu|\boldsymbol{N},\boldsymbol{\theta^0},\boldsymbol{m})$, using data, together with the 95 \% CI at the $m^{\mathrm{sig}}=1000$ GeV is drawn on Fig.~\ref{bayessol::observedul}.\par

%============================= Figure ================================
\begin{figure}[htpb]
\begin{center}
\includegraphics[width=127mm]{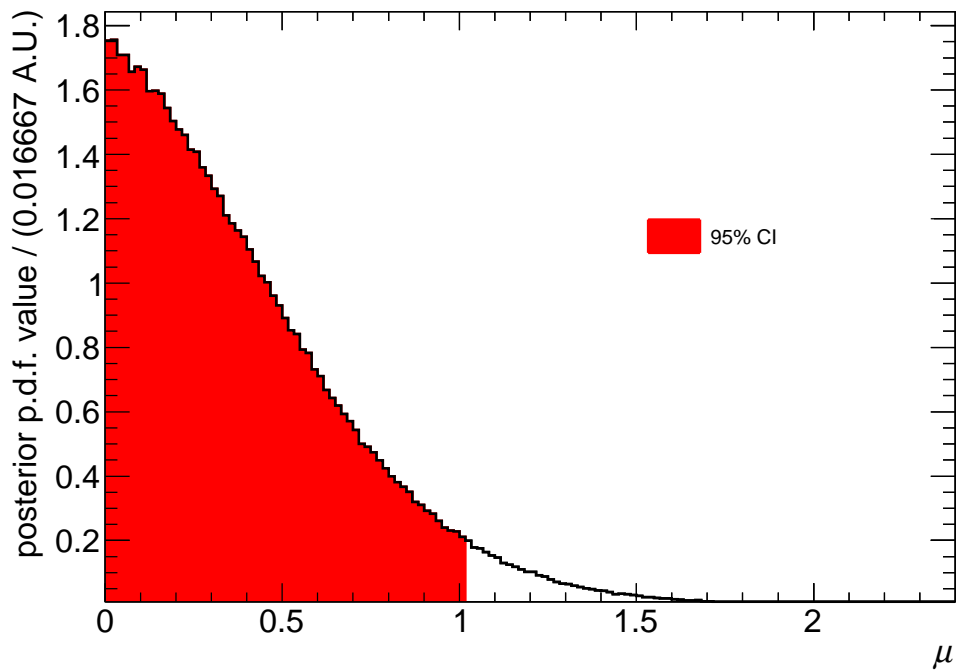} 
\vspace{-3mm}
\caption{Posterior p.d.f. for the parameter $\mu$ at $m^{\mathrm{sig}}=1000$ GeV.}
\end{center}
\labelf{bayessol::observedul}
\vspace{-5mm}
\end{figure}

The expected upper limit sampling distributions are produced for a case of only background processes presence. Unconditional $b$ ensembles of 12500 PEs are produced as it is described in Sec.~\ref{sec:freq} by means of $\mathrm{L_B}$ and $\mathrm{P_0}$, using the nuisance parameters' values from the fit of the posterior in Eq.~\ref{Eq::BayesTheorem}, given data and $\mu$ set to $0$. The sampling distribution of the Bayesian upper limit on $\mu$ from $b$ ensemble at the $m^{\mathrm{sig}}=1000$ GeV  is presented on Fig.~\ref{bayessol::samplungul}, including its observed value from Fig.~\ref{bayessol::observedul}. The conventional order statistics of the sampling distribution (median, $1\sigma$ and $2\sigma$ intervals) are also shown.\par

%============================= Figure ================================
\begin{figure}[htpb]
\begin{center}
\includegraphics[width=127mm]{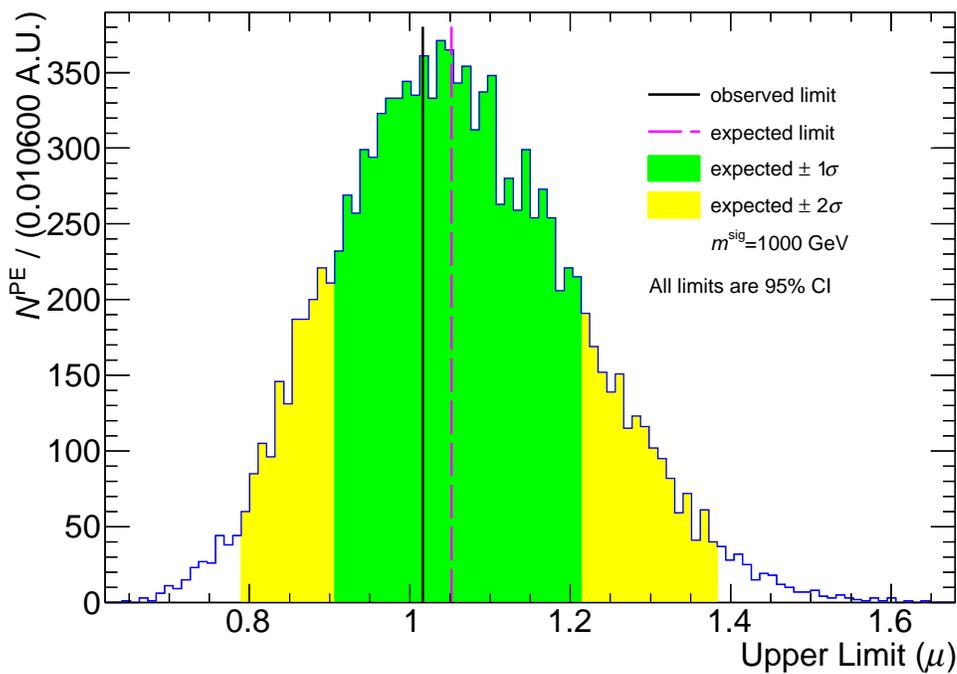} 
\vspace{-3mm}
\caption{Bayesian upper limit's $b$ ensemble and observed value at $m^{\mathrm{sig}}=1000$ GeV.}
\end{center}
\labelf{bayessol::samplungul}
\vspace{-5mm}
\end{figure}

Bayesian upper limits on $N^{\mathrm{sig}}$ as a function of $m^{\mathrm{sig}}$ are shown on Fig.~\ref{bayessol::excl}. The corresponding frequentist upper limits from Fig.~\ref{freqsol::excl} are overlaid for a comparison.\par

%============================= Figure ===============================
\begin{figure}[htpb]
\begin{center}
\includegraphics[width=127mm]{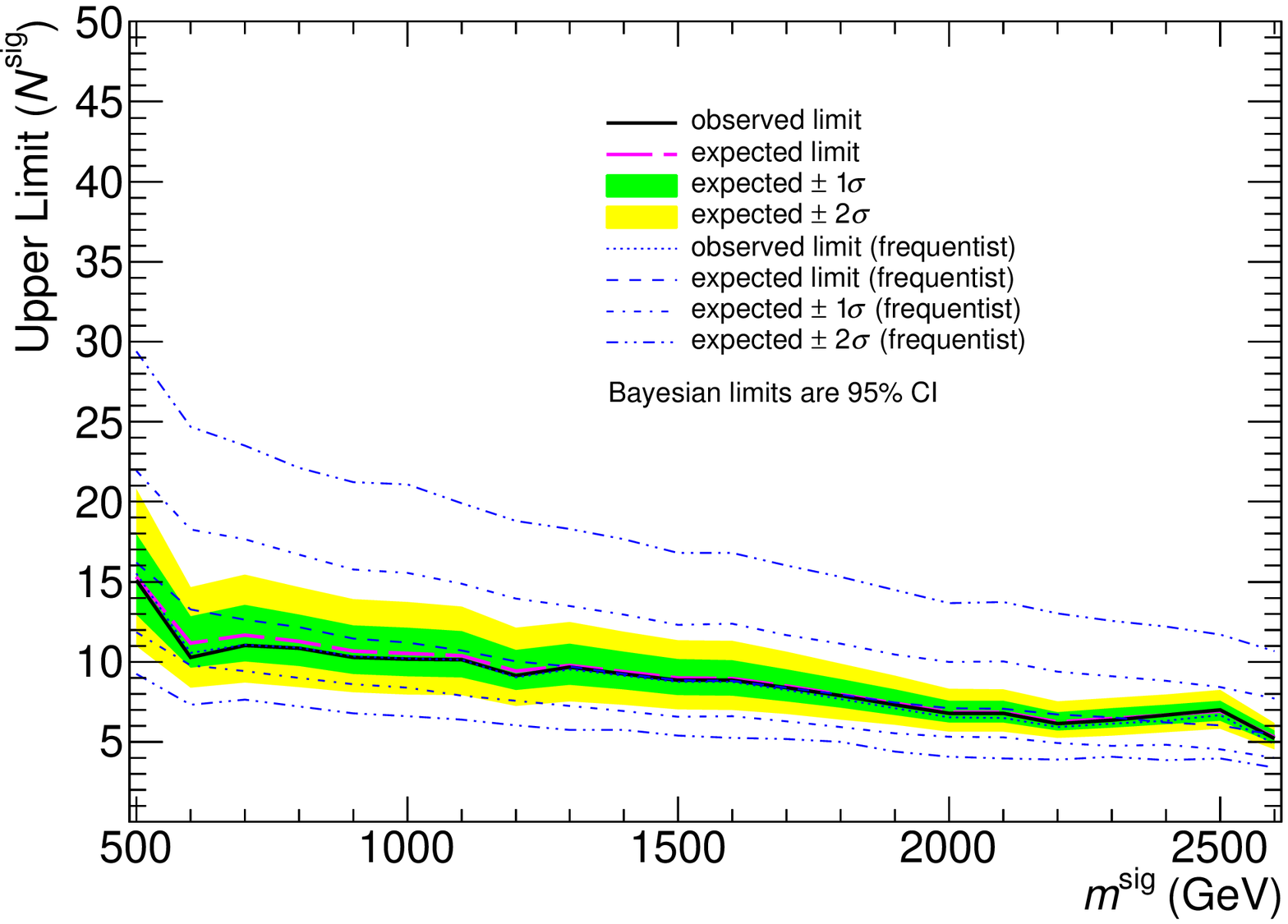} 
\vspace{-3mm}
\caption{Bayesian upper limits on $N^{\mathrm{sig}}$ as a function of $m^{\mathrm{sig}}$.}
\end{center}
\labelf{bayessol::excl}
\vspace{-5mm}
\end{figure}

As it follows from Fig.~\ref{bayessol::excl} the Bayesian and frequentist observed limit curves are compatible, and so are physics interpretations of these curves. The expected bands are diverging but the divergence is not informative: the Bayesian band is narrower than the frequentist one but in the same time it follows closer to the observed curve. A significance of deviations from the background model remains the same, varying statistical formalism: the observed curve is inside the expected $1\sigma$ band for both approaches, hence statistical and physical conclusions are compatible.

%Conclusion
\section*{Conclusion}
\label{sec:conc}

Typical experimental conditions of HEP search analyses are modeled in this work. Frequentist and Bayesian formalisms are applied to the problem and lead to the compatible statistical and physical interpretations. A choice between the approaches is proposed to be made with a scrutiny of statistical procedure's performance and reliability for each particular case, given its general complexity. It is also shown that the choice of a statistical treatment to a systematic uncertainty affects the results of a new model test and, therefore, such a treatment is recommended to be clearly described in publications of search analyses in HEP.

%Acknowledgments
\section*{Acknowledgments}
\label{sec:ackn}

The great thanks are to SRC of RF - IHEP of NRC KI Central Linux Cluster \cite{IHEP_Protvino} and SRC of RF - IHEP of NRC KI IT Department for their extensive computing support of the presented studies during whole period of the work. The particular thanks are to Alexey Myagkov and Evgeniya Cheremushkina for their criticism.

%Bibliography
\bibliographystyle{pepan}
\bibliography{StatisticalMethods}

\end{document}